\documentclass[a4paper,twoside]{article}

\usepackage{ascmac}

\usepackage{epsfig}
\usepackage{subcaption}
\usepackage{calc}
\usepackage{amssymb}
\usepackage{amstext}
\usepackage{amsmath}
\usepackage{amsthm}
\usepackage{multicol}
\usepackage{pslatex}
\usepackage{apalike}
\usepackage{algorithm2e}
\usepackage[bottom]{footmisc}
\usepackage{SCITEPRESS}     % Please add other packages that you may need BEFORE the SCITEPRESS.sty package.

\begin{document}

% \title{Software Spatial-Data Analysis\\ towards Visualization and Statistics of Internal Software Execution}

\title{Software Space Analytics: Towards Visualization and Statistics of Internal Software Execution}

\author{\authorname{Shinobu Saito\sup{1}\orcidAuthor{0000-0002-6259-3521}}
\affiliation{\sup{1}NTT, Inc., Musashino-shi, Tokyo, Japan}
\email{shinobu.saito@ntt.comm}
}

% \keywords{The paper must have at least one keyword. The text must be set to 9-point font size and without the use of bold or italic font style. For more than one keyword, please use a comma as a separator. Keywords must be titlecased.}


% \abstract{The abstract should summarize the contents of the paper and should contain at least 70 and at most 200 words. The text must be set to 9-point font size.}

\abstract{In software maintenance work, software architects and programmers need to identify modules that require modification or deletion. Whilst user requests and bug reports are utilised for this purpose, evaluating the execution status of modules within the software is also crucial. This paper, therefore, applies spatial statistics to assess internal software execution data. First, we define a software space dataset, viewing the software's internal structure as a space based on module call relationships. Then, using spatial statistics, we conduct the visualization of spatial clusters and the statistical testing using spatial measures. Finally, we consider the usefulness of spatial statistics in the software engineering domain and future challenges.}

\onecolumn \maketitle \normalsize \setcounter{footnote}{0} \vfill

\section{\uppercase{Introduction}}
\label{sec:introduction}
It was once commonly said, “Never touch a running system,” but today, the importance of modifying software in operation to meet changing user needs or removing it when no longer needed is emphasized~\cite{Nayebi2024}. Software architects and programmers utilize users' requirements changes and bug reports to identify targets for software modification or removal. In addition to the external environment of the software, understanding and evaluating the software's internal environment—such as which modules are frequently called or which modules are rarely called—provides crucial information for maintaining the existing software (e.g., module modification and removal).

This paper proposes a novel approach, \textit{Software Space Analytics}, which applies spatial statistics to evaluate the internal execution of software. Spatial statistics is a general term for statistical modeling and analysis methods targeting spatial data, which is data with location information~\cite{Cress2015}. Its application and practice have been advancing across various fields. Examples include understanding weather conditions and soil contamination, predicting disease risks and infection distributions, and investigating the habitat distributions of birds and animals.

As a technique for analyzing software execution internally, dynamic analysis collects runtime data from the programs that constitute the software and analyzes the actual behavior and performance of those programs. Due to its characteristic of analyzing the exact operation of programs, various research approaches have been proposed and implemented. A representative example is the analysis of program execution paths~\cite{Siala2024,Bergmayr2016,Briand2006,Systa2000}. These studies propose methods to visualize the behavior of programs during execution using models such as UML sequence diagrams. Additionally, numerous studies on fault localization have been proposed to identify the causes of program defects and reproduce the conditions under which they occur~\cite{Li2025,Jones2002}. Furthermore, many techniques have been proposed to discover rules and patterns (e.g., dynamic invariants) regarding the specifications of program behavior~\cite{Huang2024,Ernst2001}.

Our approach is also a form of dynamic analysis, as it analyzes software execution data to examine the software's internal state. On the other hand, the existing dynamic analysis studies mentioned above do not utilize spatial statistics. Moreover, even when considering the broader field of software engineering, research utilizing spatial statistics remains scarce. One of the few studies proposes modeling and predicting the resource allocation status for virtual machines on cloud computing using Kriging~\cite{Matheron1963}, a method within spatial statistics, and applying this to dynamic resource allocation~\cite{Gambi2012}. To the best of our knowledge, this paper is the first to apply spatial statistics to software execution data.

In contrast to existing studies that apply spatial statistics to physical space (e.g., land, soil, forests), this paper applies them to space within software. In spatial statistics, the first step is to evaluate whether the target data exhibits spatial autocorrelation. Spatial autocorrelation represents the degree of similarity (or dissimilarity) with surrounding zones. If spatial autocorrelation exists in the target data, various spatial statistical techniques become applicable. Therefore, this paper sets one research question (RQ) for applying spatial statistics to software space.

\begin{itembox}[l]{Research Question (RQ)}
% \textbf{Does software execution data exhibit spatial autocorrelation?}
Does software execution data exhibit spatial autocorrelation?
\end{itembox}

The structure of this paper is outlined below. Section 2 provides an overview of spatial statistics. Section 3 proposes our approach, software space analytics. We define a software space dataset based on the call relationships between modules within the software. This software space dataset is derived from log data collected during software execution. Section 4 presents a case study that applies spatial statistics visualization and statistical testing to this dataset, addressing the research question. We analyze two software systems actually deployed in enterprises. Section 5 discusses the case study results, and Section 6 presents the conclusion and future challenges.

\section{\uppercase{Spatial Statistics}}

\begin{figure}[t]
\begin{center}
\includegraphics[width=0.95\columnwidth]{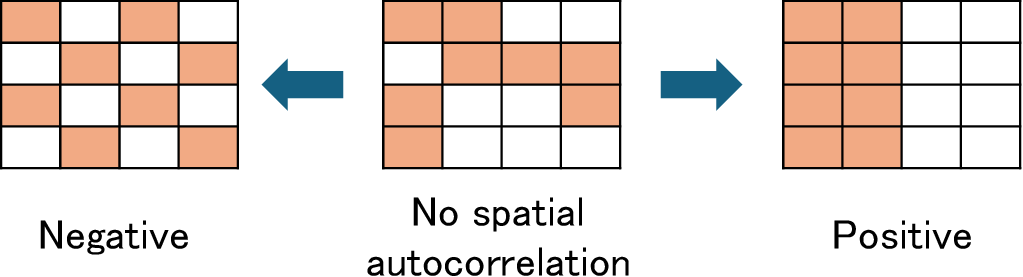}
\end{center}
\caption{Spatial autocorrelation.}
\label{fig:SA_image}
\end{figure}

One fundamental property of spatial data is spatial autocorrelation. It describes the tendency of nearby locations to have similar values (i.e., positive autocorrelation). Spatial autocorrelation is often referred to as the First Law of Geography, and its importance has long been recognized~\cite{Tobler1970,Fisher1935}.
Spatial autocorrelation can be either positive or negative. Positive autocorrelation indicates that neighborhoods tend to share similar characteristics, while negative autocorrelation suggests neighborhoods exhibit opposite tendencies. As shown in Figure~\ref{fig:SA_image}, spatial data with positive autocorrelation forms patterns where similar colors cluster together, whereas spatial data with negative autocorrelation produces checkerboard-like patterns. When there is no spatial autocorrelation, color cells appear randomly. An example of positive spatial autocorrelation is high population density areas being surrounded by other high population density areas. An example of negative spatial autocorrelation is the clustering of small stores with low sales around large stores with high sales.

\subsection{Proximity Matrix}

\begin{figure}[t]
\begin{center}
\includegraphics[width=0.95\columnwidth]{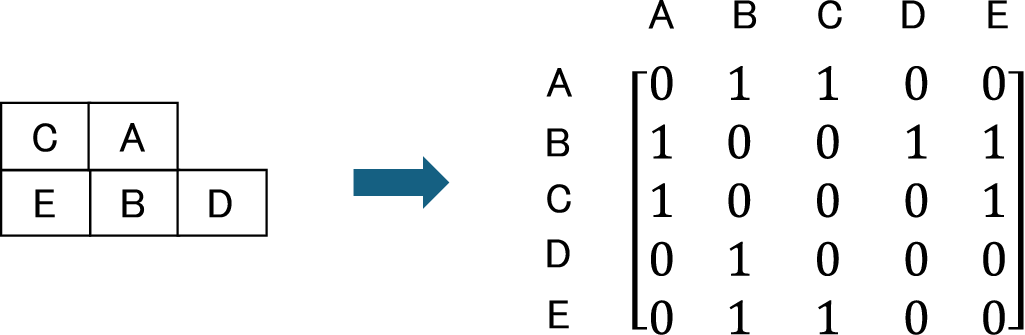}
\end{center}
\caption{Raster model (left) and its corresponding proximity matrix (rigfht).}
% \caption{Graph-based model of module dependencies (left) and proximity matrix (rigfht).}
\label{fig:Proximity_Matrix}
\end{figure}

To represent spatial autocorrelation, it is necessary to define a neighborhood. A proximity matrix is used to express the proximity relationships between zones. It is a square matrix ($N \times N$ ; $N$ is the number of zones) and a symmetric matrix. Figure~\ref{fig:Proximity_Matrix} shows the raster model and its corresponding proximity matrix. The raster model represents spatial data as a regular grid of cells. This raster model comprises five zones, designated as zones A through E. The corresponding proximity matrix ($5\times5$) contains the spatial proximity of zones A through E in the raster model.

Here, let $w_{ij}$ denote the proximity between zone $i$ and zone $j$. The proximity matrix is a matrix with the proximity $w_{ij}$ in the $i$-th row and $j$-th column. While several methods exist for defining proximity, we adopt the Rook method, which defines zones sharing boundaries as neighbors. Thus, since zone A shares boundaries with zones B and C, these two zones become its neighbors ($w_{12}=w_{21}=1,\, w_{13}=w_{31}=1$). Since spatial autocorrelation within the same zone is not considered, the diagonal elements of the proximity matrix are zero ($w_{11}=w_{22}=w_{33}=w_{44}=w_{55}=0
$).

\subsection{Spatial Statistical Measures}

Statistical measures for evaluating spatial autocorrelation fall into two categories: the Global Indicator of Spatial Association (GISA), which assesses the strength of spatial autocorrelation across the entire space, and the Local Indicator of Spatial Association (LISA), which evaluates the strength of local spatial autocorrelation within each zone~\cite{Anselin1995}. We then introduce Moran's I statistic and Local Moran's I statistic~\cite{Moran1950}, which are representative metrics for GISA and LISA, respectively.

\subsubsection{Moran's I statistic}

Moran's I statistic evaluates the spatial autocorrelation of sample data ($y_{1},...,y_{N}$) obtained from $N$ zones. The definition formula for Moran's I statistic ($I$) is given below. $\bar{y}$ denotes the sample mean.

$$
I=\frac{N}{\sum_{i}^{}\sum_{j}^{}w_{ij}}\frac{\sum_{i}^{}\sum_{j}^{}w_{ij}(y_{j}-\bar{y})(y_{i}-\bar{y})}{\sum_{i}^{}(y_{i}-\bar{y})^2}
$$

Moran's I statistic is a metric that evaluates the strength of correlation between a zone and its neighbors. It yields positive values when spatial data exhibit positive autocorrelation and negative values when negative autocorrelation is present. When the proximity matrix is row-normalized, Moran's I can theoretically reach a maximum value of 1, allowing for interpretation similar to the commonly used correlation coefficient.

\subsubsection{Local Moran's I statistic}

The representative LISA, Local Moran's I statistic ($I_{i}$), is a statistic obtained by partitioning Moran's I statistic for each zone as described below.

$$I=\frac{1}{N}\sum_{i=1}^{N}I_{i}$$

Local Moran's I statistic in zone $i$ is defined by the following formula.

$$
I_{i}=\frac{1}{m}(y_{i}-\bar{y})\sum_{j}^{}{w_{ij}(y_{j}-\bar{y})}
$$

Here, $m$ is the constant obtained during the decomposition process of Moran's I statistic ($I$).

$$
m=\frac{1}{n-1}\sum_{i=1,i\ne j}^{I}(y_{j}-\bar{y})^2-\bar{y}^2
$$

\subsection{Statistical Test for Local Moran's I}

Using Local Moran's I statistic allows for a statistical test to determine whether autocorrelation exists in each zone. This test utilizes the $z$ value obtained under the null hypothesis that “there is no spatial autocorrelation around zone $i$.”

$$
z[I_{i}]=\frac{I_{i}-E[I_{i}]}{\sqrt{Var[I_{i}]}}
$$

$E[I_{i}]$ and $Var[I_{i}]$ are the expected value and variance under the null hypothesis.

\subsection{Spatial Clusters}

In spatial statistics, Moran's scatterplot is known for visualizing the distribution of spatial autocorrelation measured by Local Moran's I statistic. Based on the definition of Local Moran's I statistic, each zone is classified into one of four categories as follows, based on the combination of positive and negative values of $y_{i}-\bar{y}$ and $\sum_{j}^{}w_{ij}(y_{j}-\bar{y})$. These four categoriess are referred to as spatial clusters: hot spot, cool spot, high-value outlier, and low-value outlier.

\begin{itemize}
 \item \textbf{Hot spot (High-High)}
    \begin{itemize}
    \item $y_{i}-\bar{y}>0\land\sum_{j}^{}w_{ij}(y_{j}-\bar{y})>0$
    \item High-value zone surrounded by high-values
    \end{itemize}    
 \item \textbf{Cool spot (Low-Low)}
    \begin{itemize}
    \item $y_{i}-\bar{y}<0\land\sum_{j}^{}w_{ij}(y_{j}-\bar{y})<0$
    \item Low-value zone surrounded by low-values
    \end{itemize}
 \item \textbf{High-value outlier (High-Low)}
    \begin{itemize}
    \item $y_{i}-\bar{y}>0\land\sum_{j}^{}w_{ij}(y_{j}-\bar{y})<0$
    \item High-value zone surrounded by low-values
    \end{itemize}
 \item \textbf{Low-value outlier (Low-High)}
    \begin{itemize}
    \item $y_{i}-\bar{y}<0\land\sum_{j}^{}w_{ij}(y_{j}-\bar{y})>0$
    \item Low-value zone surrounded by high-values
    \end{itemize}
\end{itemize}

\begin{figure}[t]
\begin{center}

\includegraphics[width=\columnwidth]{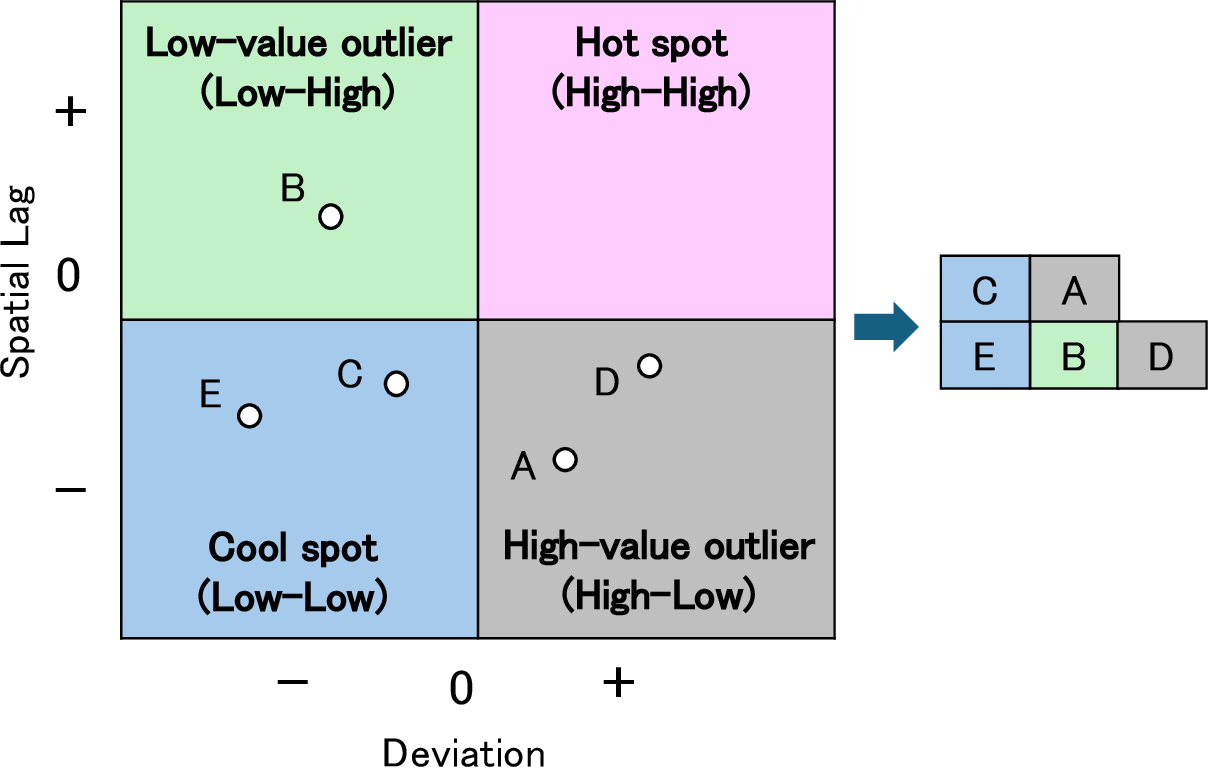}
\end{center}
\caption{Moran scatter plot (left) and visualization of raster model based on spatial clusters (right).}
\label{fig:SP_Zone}
\end{figure}

\subsection{Moran's Scatter Plot}
Setting $y$ as follows produces a Moran's scatter plot for $y$ that appears as shown on the left side of Figure~\ref{fig:SP_Zone}. 
$$
y=[y_{A}, y_{B}, y_{C},y_{D},y_{E}]=[95, 55, 60, 83, 32]
$$

To represent spatial clusters, the four quadrants are colored with four distinct colors: red (hot spot), blue (cool spot), gray (high-value outlier), and green (low-value outlier). Then, on the right side of the figure, the five zones of the raster model are colored in three different colors (i.e., gray, green, and blue) based on Moran's scatter plot.

\begin{figure*}[t]
\begin{center}

\includegraphics[width=1.9\columnwidth]{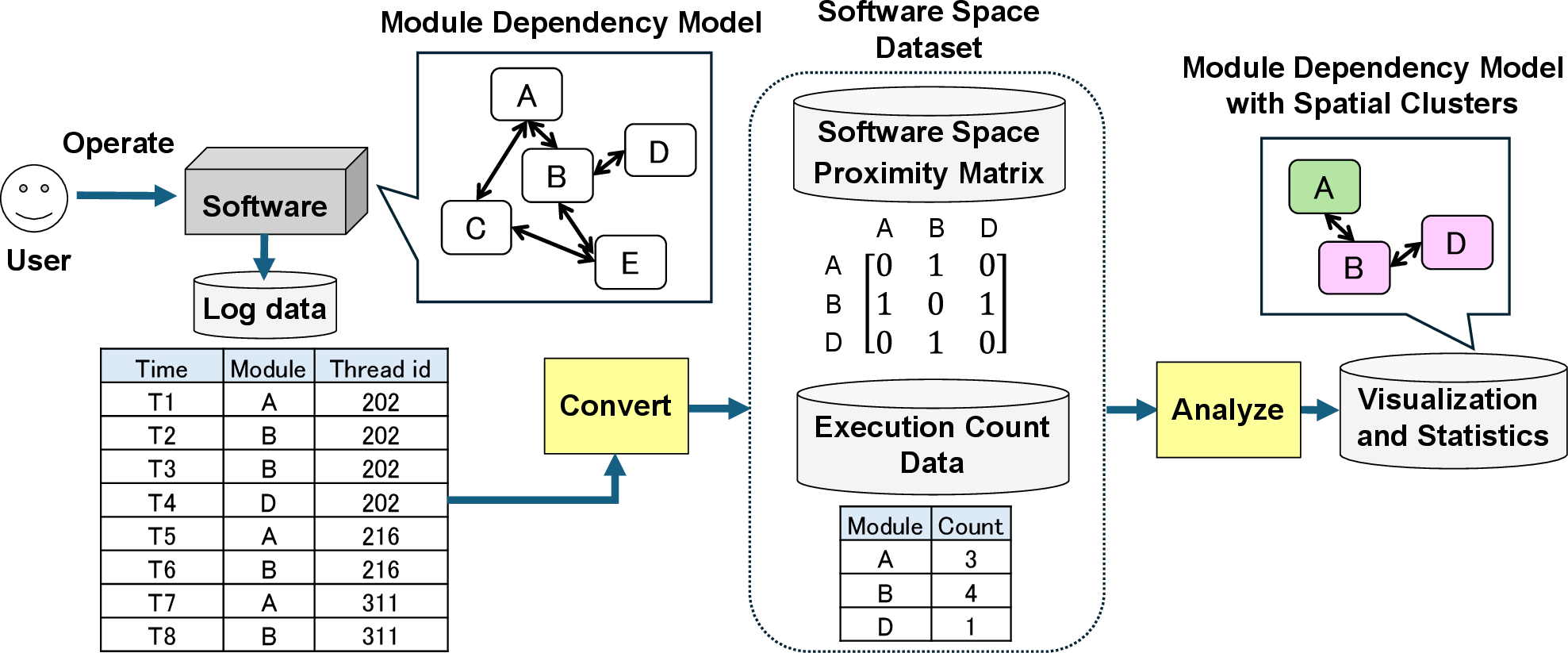}
\end{center}

\caption{Schematic of Software Space Analytics.}
\label{fig:Approach}
\end{figure*}

\section{\uppercase{Software Space Analytics}}

Our approach, \textit{Software Space Analytics}, uses spatial statistics to evaluate the properties and relationships of module execution within software. Figure~\ref{fig:Approach} shows a schematic of software space analytics. As users on the far left operate the software, execution log data accumulates. Transforming this log data generates a dataset for spatial statistics, which we refer to as the \textit{Software Space Dataset}. It consists of two components: software space proximity matrix and execution count data. We then apply visualization and statistical techniques used in spatial statistics to the dataset.

\subsection{Software Space Proximity Matrix}

The lower left of Figure~\ref{fig:Approach} shows log data obtained from software execution. While log data formats vary, we focus on three elements always obtainable from log data: Time, Module, and Thread id. ``Time'' is the time when the module's execution started. ``Module'' is the name identifying a function within the software (e.g., class name). ``Thread id'' is an identifier assigned to a specific set of software execution traces. The log data in the lower left of the figure holds eight log records, consisting of three execution traces. The records from the first to the fourth line are execution traces for thread ID ``202''. Similarly, the records from the fifth to the sixth line are execution traces for thread ID ``216'', and the records from the seventh to the eighth line are execution traces for thread ID ``311''.

From the log data, we create a software space proximity matrix. Here, the internal modules of the software correspond to each zone in the proximity matrix. To identify the neighborhood of each module, we extract function call relationships from this log data. In this sample, two call relationships exist: from A to B and from B to D. Therefore, the elements in the respective related rows and columns are set to ``1''. Since the proximity matrix is symmetric, a single call relationship is represented by two elements.
In the log data, modules D and A are executing in lines 4 and 5. There is no function call relationship from D to A because these two records belong to different execution traces. While B is executed twice consecutively in the log data, diagonal elements in the proximity matrix are set to ``0''. Therefore, the element in the 2nd row and 2nd column is set to ``0''. Through this transformation, a 3-row, 3-column software space proximity matrix is created from the log data.

\subsection{Execution Count Data}

Execution count data stores the number of times each module was executed in the ``Count'' column. For example, based on the log data in the lower left of Figure~\ref{fig:Approach}, modules A, B, and D were executed a total of 3, 4, and 1 times, respectively. As shown in the upper left of the figure, this software consists of five modules in total (A through E). However, referring to the log data, since modules C and E were not executed, the execution count data contains only three records corresponding to the three executed modules (i.e., A, B, and D). Since each module corresponds to a zone in spatial statistics, the execution count values represent the sample data obtained in each zone.

\vskip\baselineskip

By converting the log data to the software space dataset as described above, various visualization and statistical analysis methods in spatial statistics can be applied to the log data accumulated during software operation.

\begin{figure*}[t]
\begin{center}
\includegraphics[width=2.0\columnwidth]{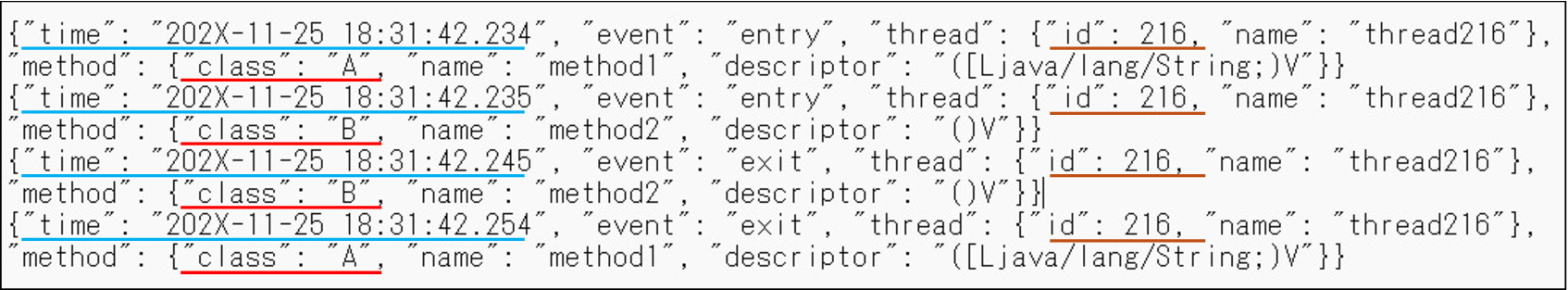}
\end{center}
\caption{Execution log data (JSON format).}
\label{fig:log_JSON}
\end{figure*}

\section{\uppercase{Case Study}}

This chapter reports a case study in which we collected execution log data from two software systems in actual operation at a company and performed software space analysis. Table \ref{table:Dataset_overview} shows the overview of two datasets obtained from two software systems. 

\begin{table}[b]
  \caption{Dataset Overview.}
  \label{table:Dataset_overview}
\centering
\small

  \begin{tabular}{p{29mm}rr}
\hline
&\textbf{Dataset 1}&\textbf{Dataset 2}\\\hline
Software&Apache Ofbiz&Lavagna\\
Version &18.12&1.1.9\\
Collection period&2 months&2 months\\
No. of users&3 operators&3 engineers\\\hline

No. of total classes&1,137 &256 \\
No. of classes called by other classes&247 &65 \\\hline

\end{tabular}
\end{table}

 % without inner classes

\subsection{Software and Users}

Both pieces of software in operation at the company discussed in this case are open-source software. Apache OFBiz\footnote{https://github.com/apache/ofbiz-framework} is a comprehensive business application suite that encompasses various functions, including accounting, asset management, human resources, content management, and customer relationship management. Lavagna\footnote{https://lavagna.io/} is a project management tool for small teams. It provides standard features for general project management, such as task assignment and progress tracking, and also allows customization of Kanban boards. These two software systems were operated over two months, with three users (i.e., operators and developers) each continuously performing operations in their actual work environments.

\subsection{Data Collection}

To collect log data, we used JavaAgent\footnote{https://docs.oracle.com/javase/8/docs/api/java/lang/\\instrument/package-summary.html}. JavaAgent weaves logging code into the bytecode at the moment the target software's Java files are loaded into the virtual machine. So, we did not need to modify the target software's program to collect execution log data.

Using JavaAgent, we were able to collect log data detailing when Java methods within the software started (entry) and ended (exit). Figure~\ref{fig:log_JSON} shows the contents of this log data (partially masked). It consists of four records representing the execution trace of thread ID ``216''. The first line indicates the start of method 1 in class A. The second line shows the start of method 2 in class B, and the third line shows the end of method 2. The final fourth line shows the end of method 1 in class A. From this, we can see that these four records represent log data showing class 1 calling class 2. In this way, we identified class-level call relationships from the log data.

Table \ref{table:Dataset_overview} lists the total number of classes, excluding inner classes, obtained through static analysis of the program and the total number of executing classes obtained from log data. For both datasets, the number of executed classes is less than 30\% of the total number of classes.

\begin{figure*}[t]

\small
  \begin{minipage}[b]{0.49\linewidth}
    \centering
    \includegraphics[keepaspectratio, width=0.85\columnwidth]{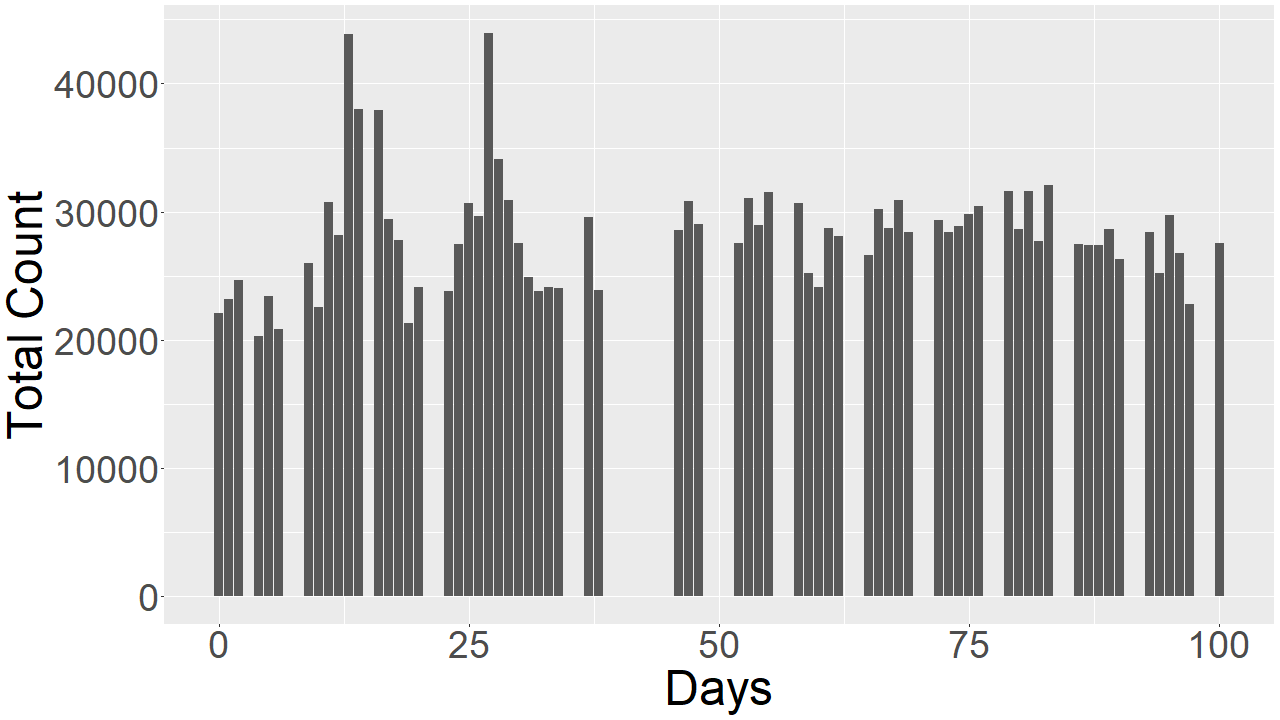}
    \caption{Time-series trend in execution count (Dataset 1)}
    \label{fig:CountData1}
  \end{minipage}
  \begin{minipage}[b]{0.49\linewidth}
    \centering
    \includegraphics[keepaspectratio, width=0.85\columnwidth]{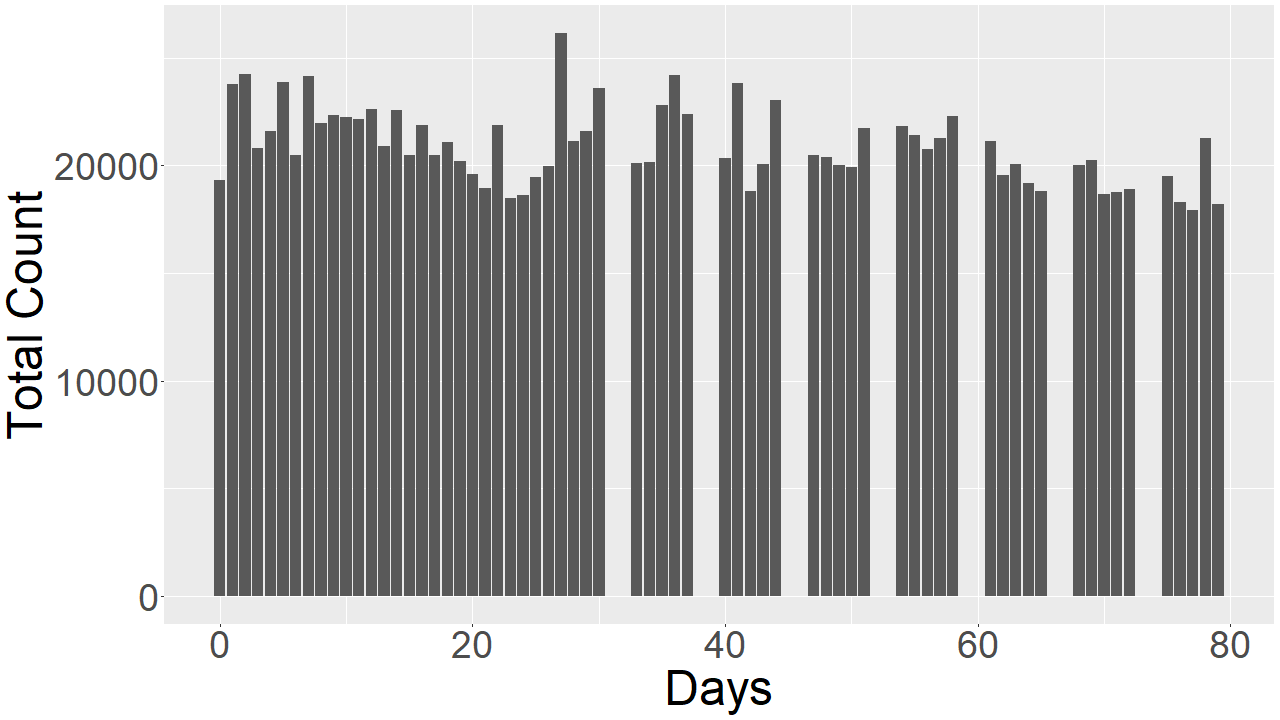}
    \caption{Time-series trend in execution count (Dataset 2)}
    \label{fig:CountData2}
  \end{minipage}

  \begin{minipage}[b]{0.49\linewidth}
    \centering
    \includegraphics[keepaspectratio, width=0.92\columnwidth]{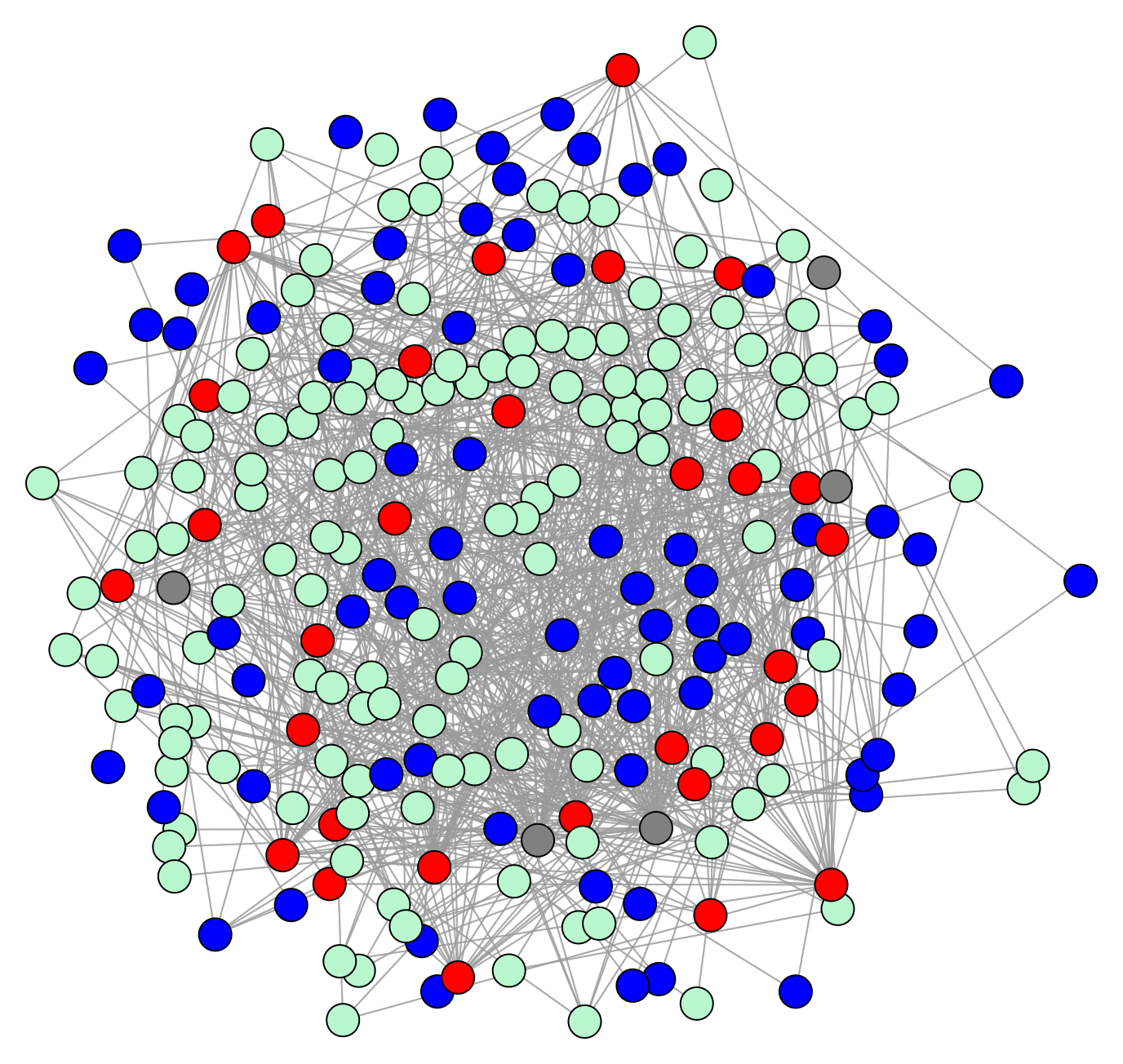}
    \includegraphics[keepaspectratio, width=0.8\columnwidth]{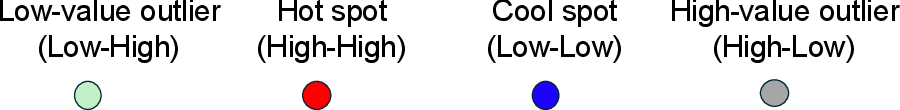}
    \caption{Module dependency model with spatial clusters\\(Dataset 1)}
    \label{fig:HC_1}
  \end{minipage}
  \begin{minipage}[b]{0.49\linewidth}
    \centering
    \includegraphics[keepaspectratio, width=0.85\columnwidth]{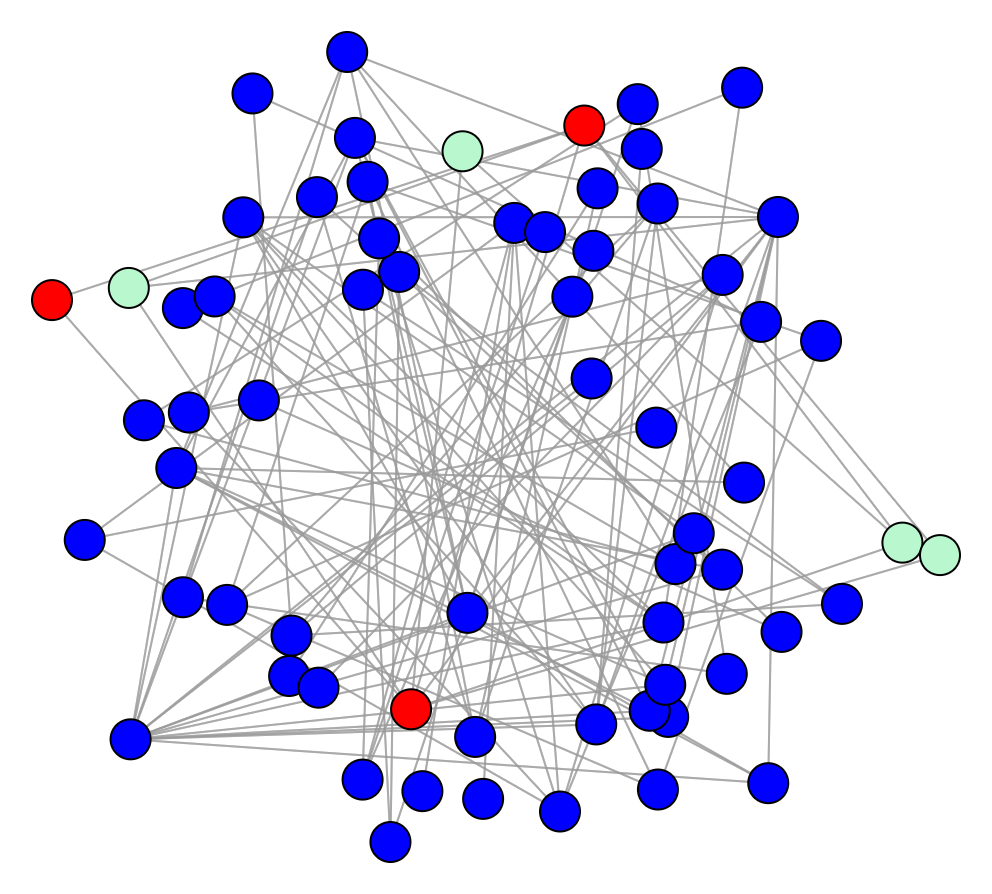}
    \includegraphics[keepaspectratio, width=0.8\columnwidth]{Figs/legend.eps}
    \caption{Module dependency model with spatial clusters\\(Dataset 2)}
    \label{fig:HC_2}
  \end{minipage}

\end{figure*}

Figure~\ref{fig:CountData1} and~\ref{fig:CountData2} show the time-series trend of execution counts for data collected from the two software applications. The vertical axis represents the total execution count for all modules (i.e., Java classes) within the software, while the horizontal axis denotes days. During almost all weekdays of the operational period, both software recorded similar execution counts. Due to corporate operations, weekends and long holidays typically represent periods with minimal user activity.

\subsection{Visualization of Spatial Clusters}

We converted the software space dataset from collected execution log data. We then calculated Local Moran's I statistic for each module (i.e., Java class), and visualized the calculation results. Figures~\ref{fig:HC_1} and~\ref{fig:HC_2} are visualizations of the graph-based module dependency model based on the values of Local Moran's I statistic. In this model, circular objects correspond to software modules. Lines connecting circular objects indicate module call relationships.

All connected circular objects represent a neighborhood relationship. In spatial statistics, zones sharing boundaries were considered neighbors. Therefore, the number of shared boundaries (i.e., the number of neighbors) is limited by the physical boundary length of the zone, such as a block of land. In contrast, in software, a single module can theoretically call any number of other modules. So, a module called the god object~\cite{Brown1999} can exist, possessing such a large number of neighbors.

\begin{table}[b]
  \caption{Resutls of spatial cluster analysis.}
  \label{table:SP_results}
\centering
\small
  % \begin{tabular}{p{35mm}p{19mm}p{19mm}}
  \begin{tabular}{lrr}
\hline
\textbf{Cluster type (color)}&\textbf{Dataset 1}&\textbf{Dataset 2}\\\hline
Hot spot (red)&32 (13\%)&3 (5\%)\\
Cool spot (blue)&76 (31\%)&58 (89\%)\\
High-value outlier (gray)&5 (2\%)&0 (0\%)\\
Low-value outlier (green)&76 (31\%)&4 (6\%)\\\cline{2-3}
\multicolumn{1}{r}{No. of zones} &247 (100\%)&65 (100\%)\\\hline
% \multicolumn{1}{r}{No. of zones}& 306 (100\%)& 78 (100\%)
\end{tabular}
\end{table}

 \begin{figure*}[t]
  \begin{minipage}[b]{0.50\linewidth}
    \centering
    \includegraphics[keepaspectratio, width=0.92\columnwidth]{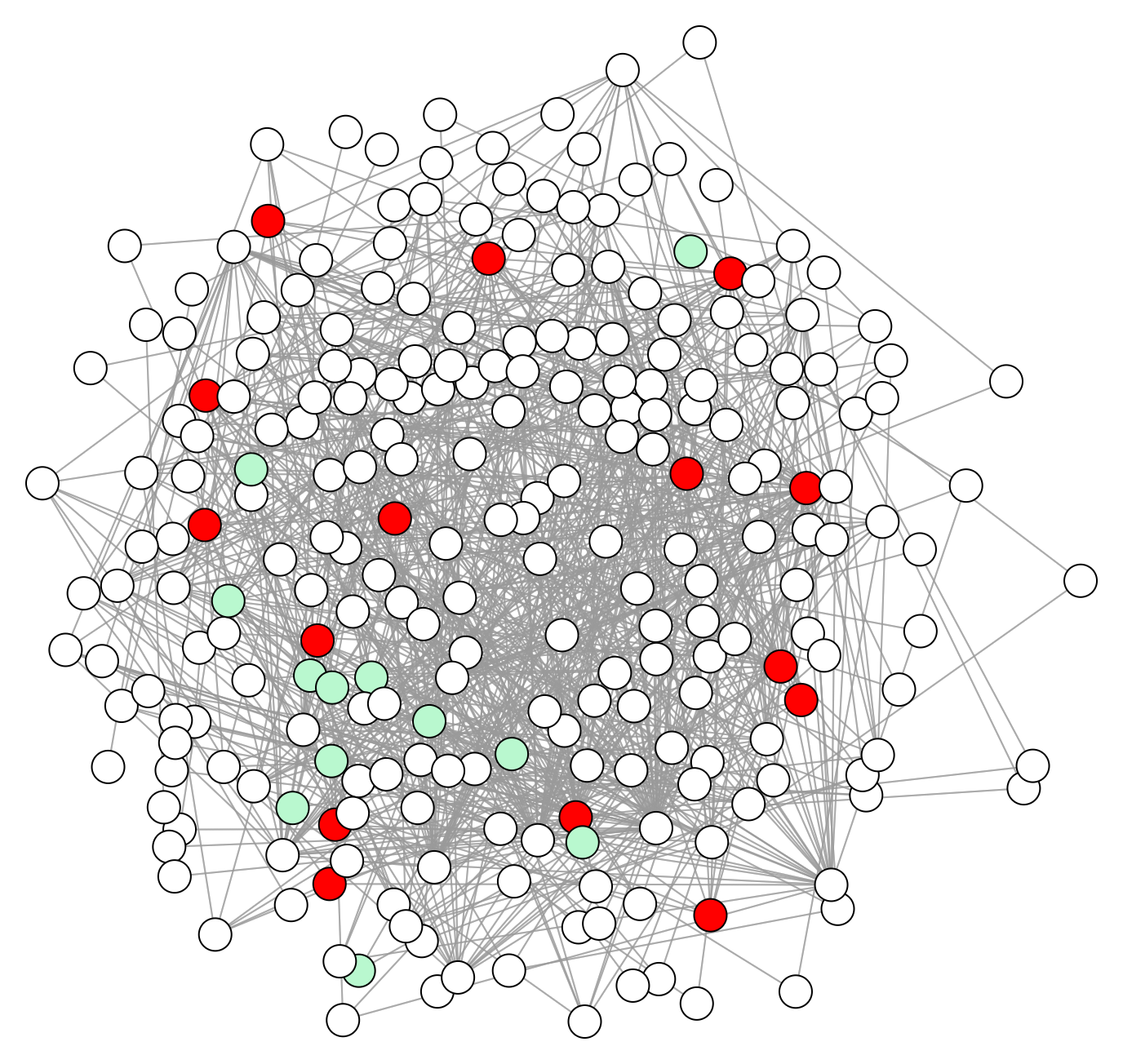}
    \includegraphics[keepaspectratio, width=0.8\columnwidth]{Figs/legend.eps}
    \caption{Module dependency model with only statistically\\ significant zones colored (Dataset 1)}
    \label{fig:P_1}
  \end{minipage}
  \begin{minipage}[b]{0.50\linewidth}
    \centering
    \includegraphics[keepaspectratio, width=0.85\columnwidth]{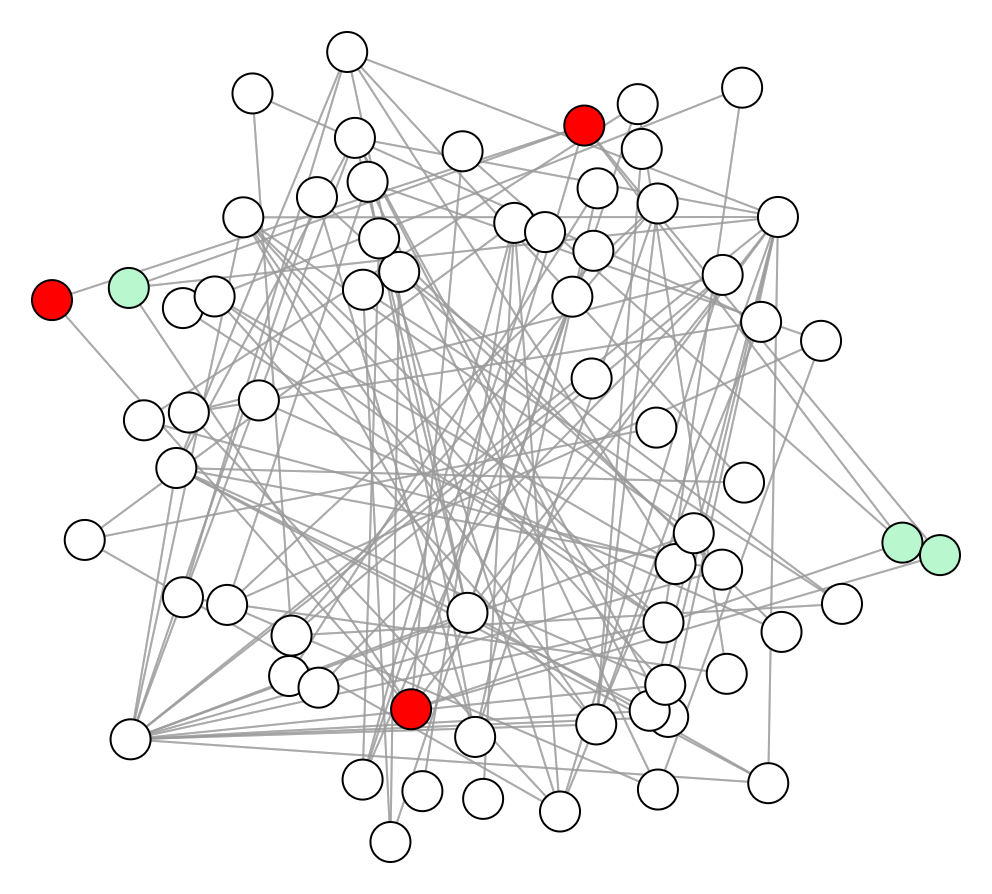}
    \includegraphics[keepaspectratio, width=0.8\columnwidth]{Figs/legend.eps}
    \caption{Module dependency model with only statistically\\ significant zones colored (Dataset 2)}
    \label{fig:P_2}
  \end{minipage}
\end{figure*}

Table \ref{table:SP_results} describes the results of the spatial clusters analysis for the two software space datasets. Dataset 1 contains numerous modules classified as low-value outlier (green) and cool spot (blue), followed by hot spot (red). In contrast, Dataset 2 contains many modules belonging to cool spot, with only a few low-value outlier and hot spot modules observed. Furthermore, modules classified as high-value outlier (gray) are scarcely seen in either dataset.

\begin{table}[b]
  \caption{Statistical test resutls for each spatial cluster.}
  \label{table:p-value_results}
\centering
\small
  % \begin{tabular}{p{35mm}p{19mm}p{19mm}}
  \begin{tabular}{lrr}
\hline
&\multicolumn{2}{c}{\textbf{Statistically Significant Zones}}\\
&\multicolumn{2}{c}{\textbf{$P_{r}\{z \neq E(I_{i}) \} \le 0.05$}}\\\cline{2-3}
\textbf{Cluster type}&\textbf{Dataset 1}&\textbf{Dataset 2}\\\hline
Hot spot &15&3\\
Cool spot &0&0\\
High-value outlier &0&0\\
Low-value outlier &12&3\\\cline{2-3}
\multicolumn{1}{r}{Sum} &27 (11\%)&6 (9\%)\\
\multicolumn{1}{r}{No. of zones}& 247 (100\%)& 65 (100\%)\\\hline
\end{tabular}
\end{table}

\subsection{Statistical Test for Local Moran's I}

We applied the statistical test for Local Moran's I to the two datasets. The test results are described in Table~\ref{table:p-value_results}. In datasets 1 and 2, there were 27 and 6 statistically significant zones, respectively. The proportion of these significant zones within the total number of zones was approximately 11\% (=27/247) in Dataset 1,
and 9\% (=6/65) in Dataset 2.

Figures~\ref{fig:P_1} and~\ref{fig:P_2} depict a graph-based module dependency model that visualizes only modules where Local Moran's I statistic is statistically significant. In both software, statistically significant spatial autocorrelations occur solely with two types of clusters: hot spots and low-value outliers. The two types occur in nearly equal proportions. Thus, from the evaluation results of local spatial statistics, we confirmed that the spatial autocorrelations in the software execution data of the two software systems—namely, the proportion of statistically significant modules and the proportion of spatial clusters within those modules—exhibit similar trends.

\begin{itembox}[l]{Answer to RQ}
   In both software execution data, statistically significant spatial autocorrelations were observed in around 10\% of the total number of modules being called. All of them were classified as hot spot or low-value outlier.
\end{itembox}

% \textbf{Answer to RQ} AAA.

\section{\uppercase{Discussion}}
\label{sec:discussion}

\subsection{Utilization of Spatial Clusters}
Through the software space analytics, we can identify spatial clusters of modules within software. For example, modules belonging to hot spots or cool spots can be regarded as groups of modules exhibiting similar execution frequency tendencies. Conversely, modules belonging to high-value outliers or low-value outliers can be regarded as groups of modules used differently compared to surrounding modules.

Such module characteristics provide valuable information for software maintainers. For instance, hot spot modules are frequently called and thus prone to becoming performance bottlenecks. Therefore, when performance issues arise in software, investigating and modifying the hot spot module group is considered an effective solution. Modifying frequently called modules (e.g., hotspot) often yields a greater overall performance improvement than modifying slow-processing modules.
Additionally, a high-value outlier module is likely a class where responsibilities are concentrated, potentially forming a god object which not only reduces readability but also increases modification cost. Therefore, refactoring, such as redesigning responsibilities, can be prioritized.

Modules classified as cold spots or low-value outliers require a reevaluation of their roles. For example, within a group of cold spot modules, if the processing of each module is too simple and the modules are tightly coupled with each other, there is little value in keeping them split into multiple modules. It is preferable to consolidate such groups of modules. Refactoring to simplify this inter-module structure can enhance the software's maintainability. Regarding low-value outlier modules, these may handle exceptions. Software maintainers can leverage an understanding of the error-handling context to reassess the responsibilities of these modules and refine test cases for exception handling.

\subsection{Creation of Proximity Matrix}

We created a software space proximity matrix based on the content of execution log data. This means that the functions (i.e, Java classes) comprising the proximity matrix are limited to those executed at least once during the data collection period. On the other hand, the Java classes actually used in either software constitute less than 30\% of the total number of classes composing the respective software. In other words, the analysis of the software space dataset in this study could only evaluate less than 30\% of the space when viewed from the perspective of the entire software. To expand the analyzed software space, it is conceivable to consider the source code of the target software in addition to the execution log data.

We can also create a software space proximity matrix based on the call relationships of all functions within the source code through static analysis of the software source code. While this matrix can represent large-scale software spaces, it becomes a sparse matrix with many zeros. Consequently, spatial autocorrelations may be underestimated in statistical analysis, requiring caution when understanding the execution state within the software.

\section{\uppercase{Concluding Remarks}}
\label{sec:conclusion}

In this paper, we applied spatial statistics to software execution data. Specifically, we conducted the visualization of spatial clusters and the statistical testing of Local Moran's I on the two datasets. The test results from the case study indicated that statistically significant spatial autocorrelation was confirmed in approximately 10\% of all executed modules. We then discussed the usefulness of treating modules as groups (i.e., spatial clusters) and understanding their characteristics, such as hot spots, in software maintenance.

As spatial statistics encompasses not only the visualization and testing techniques, but also prediction techniques, we plan to apply them to predict the execution count of software modules. In this way, we will extend our approach to apply spatial statistics to various problems in software engineering.

\section*{\uppercase{Acknowledgements}}
The author thanks the collaborators: M. Oda, T. Shiraki, Piecemeal Technology Inc.; S. Sumita, NTT, Inc.

\bibliographystyle{apalike}
{\small
\bibliography{references}}

@article{Tobler1970,
	author =	"Tobler, W. R. ",
	title =		"A Computer Movie Simulating Urban Growth in the Detroit Region",
	journal =	"Economic Geography",
	volume =	46,
	pages =		"234--240",
	year =		1970}

@book{Fisher1935,
	author =	"Fisher, R.A.",
	title =		"The Design Of Experiments",
	publisher =	"Oliver and Boyd",
	year =		1935}

@article{Moran1950,
    author = {Moran, P. A. P.},
    title = {NOTES ON CONTINUOUS STOCHASTIC PHENOMENA},
    journal = {Biometrika},
    volume = {37},
    number = {1-2},
    pages = {17-23},
    year = {1950},
    month = {06},
    issn = {0006-3444},
    doi = {10.1093/biomet/37.1-2.17},
    url = {https://doi.org/10.1093/biomet/37.1-2.17},
    eprint = {https://academic.oup.com/biomet/article-pdf/37/1-2/17/487420/37-1-2-17.pdf},
}

@article{Nayebi2024,
    author = {Nayebi, M. and Kuznetsov, K. and Zeller, A. and Ruhe, G.},
    title = {Recommending and release planning of user-driven functionality deletion for mobile apps},
    journal = {Requirements Eng},
    volume = {29},
    number = {},
    pages = {459–480},
    year = {2024},
    month = {},
    issn = {},
    doi = {https://doi.org/10.1007/s00766-024-00430-5},
    url = {},
    eprint = {},
}

@BOOK{Cress2015,
  AUTHOR =       "Cressie, Noel",
  TITLE =        "Statistics for Spatial Data",
  PUBLISHER =    "Wiley",
  YEAR =         "2015",
  address =      "",
  edition =      ""
}

@article{Matheron1963,
    author = {Matheron, G},
    title = {Principles of geostatistics},
    journal = {Economic Geology},
    volume = {58},
    number = {8},
    pages = {1246-1266},
    year = {1963},
    month = {},
    issn = {},
    doi = {https://doi.org/10.2113/gsecongeo.58.8.1246},
    url = {},
    eprint = {},
}

@inproceedings{Gambi2012,
author = {Gambi, A. and Toffetti, G.},
title = {Modeling cloud performance with kriging},
year = {2012},
isbn = {9781467310673},
publisher = {IEEE Press},
booktitle = {Proceedings of the 34th International Conference on Software Engineering},
pages = {1439–1440},
numpages = {2},
location = {Zurich, Switzerland},
series = {ICSE '12}
}

@book{Brown1999,
author = {Brown, William H. and Malveau, Raphael C. and McCormick, Hays W. "Skip" and Mowbray, Thomas J.},
title = {AntiPatterns: Refactoring Software, Architectures, and Projects in Crisis},
year = {1998},
isbn = {0471197130},
publisher = {John Wiley \& Sons, Inc.},
address = {USA},
edition = {1st},
}

@ARTICLE{Anselin1995,
  author={Anselin, L.},
  journal={Geographical Analysis}, 
  title={Local Indicators of Spatial Association—LISA}, 
  year={1995},
  volume={27},
  number={2},
  pages={93-115},
  doi={10.1109/TSC.2015.2389236}}

@INPROCEEDINGS{Systa2000,
  author={Systa, T.},
  booktitle={Proceedings Seventh Working Conference on Reverse Engineering}, 
  title={Understanding the behavior of Java programs}, 
  year={2000},
  volume={},
  number={},
  pages={214-223},
  doi={10.1109/WCRE.2000.891472}}

@ARTICLE{Briand2006,
  author={Briand, L.C. and Labiche, Y. and Leduc, J.},
  journal={IEEE Transactions on Software Engineering}, 
  title={Toward the Reverse Engineering of UML Sequence Diagrams for Distributed Java Software}, 
  year={2006},
  volume={32},
  number={9},
  pages={642-663},
  doi={10.1109/TSE.2006.96}}

@ARTICLE{Siala2024,
  author={Siala, Hanan Abdulwahab and Lano, Kevin and Alfraihi, Hessa},
  journal={IEEE Access}, 
  title={Model-Driven Approaches for Reverse Engineering—A Systematic Literature Review}, 
  year={2024},
  volume={12},
  number={},
  pages={62558-62580},
  doi={10.1109/ACCESS.2024.3394732}}

@INPROCEEDINGS{Bergmayr2016,
  author={Bergmayr, Alexander and Bruneliere, Hugo and Cabot, Jordi and Garcia, Jokin and Mayerhofer, Tanja and Wimmer, Manuel},
  booktitle={2016 IEEE/ACM 8th International Workshop on Modeling in Software Engineering (MiSE)}, 
  title={fREX: fUML-based Reverse Engineering of Executable Behavior for Software Dynamic Analysis}, 
  year={2016},
  volume={},
  number={},
  pages={20-26},
  doi={10.1145/2896982.2896984}}

@INPROCEEDINGS{Jones2002,
  author={Jones, J.A. and Harrold, M.J. and Stasko, J.},
  booktitle={Proceedings of the 24th International Conference on Software Engineering. ICSE 2002}, 
  title={Visualization of test information to assist fault localization}, 
  year={2002},
  volume={},
  number={},
  pages={467-477},
  doi={10.1145/581396.581397}}

@INPROCEEDINGS{Li2025,
  author={Li, Chun and Li, Hui and Li, Zhong and Pan, Minxue and Li, Xuandong},
  booktitle={2025 IEEE/ACM 47th International Conference on Software Engineering (ICSE)}, 
  title={Enhancing Fault Localization in Industrial Software Systems via Contrastive Learning}, 
  year={2025},
  volume={},
  number={},
  pages={691-703},
  doi={10.1109/ICSE55347.2025.00009}}

@ARTICLE{Ernst2001,
  author={Ernst, M.D. and Cockrell, J. and Griswold, W.G. and Notkin, D.},
  journal={IEEE Transactions on Software Engineering}, 
  title={Dynamically discovering likely program invariants to support program evolution}, 
  year={2001},
  volume={27},
  number={2},
  pages={99-123},
  doi={10.1109/32.908957}}

@INPROCEEDINGS{Huang2024,
  author={Huang, Zunchen and Ravi, Srivatsan and Wang, Chao},
  booktitle={2024 39th IEEE/ACM International Conference on Automated Software Engineering (ASE)}, 
  title={Discovering Likely Program Invariants for Persistent Memory}, 
  year={2024},
  volume={},
  number={},
  pages={1795-1807},
  doi={}}

\section*{APPENDIX}
\subsubsection*{Software Space Proximity Matrix (Dataset 1)}
\begin{center}
\includegraphics[width=1.58\columnwidth]{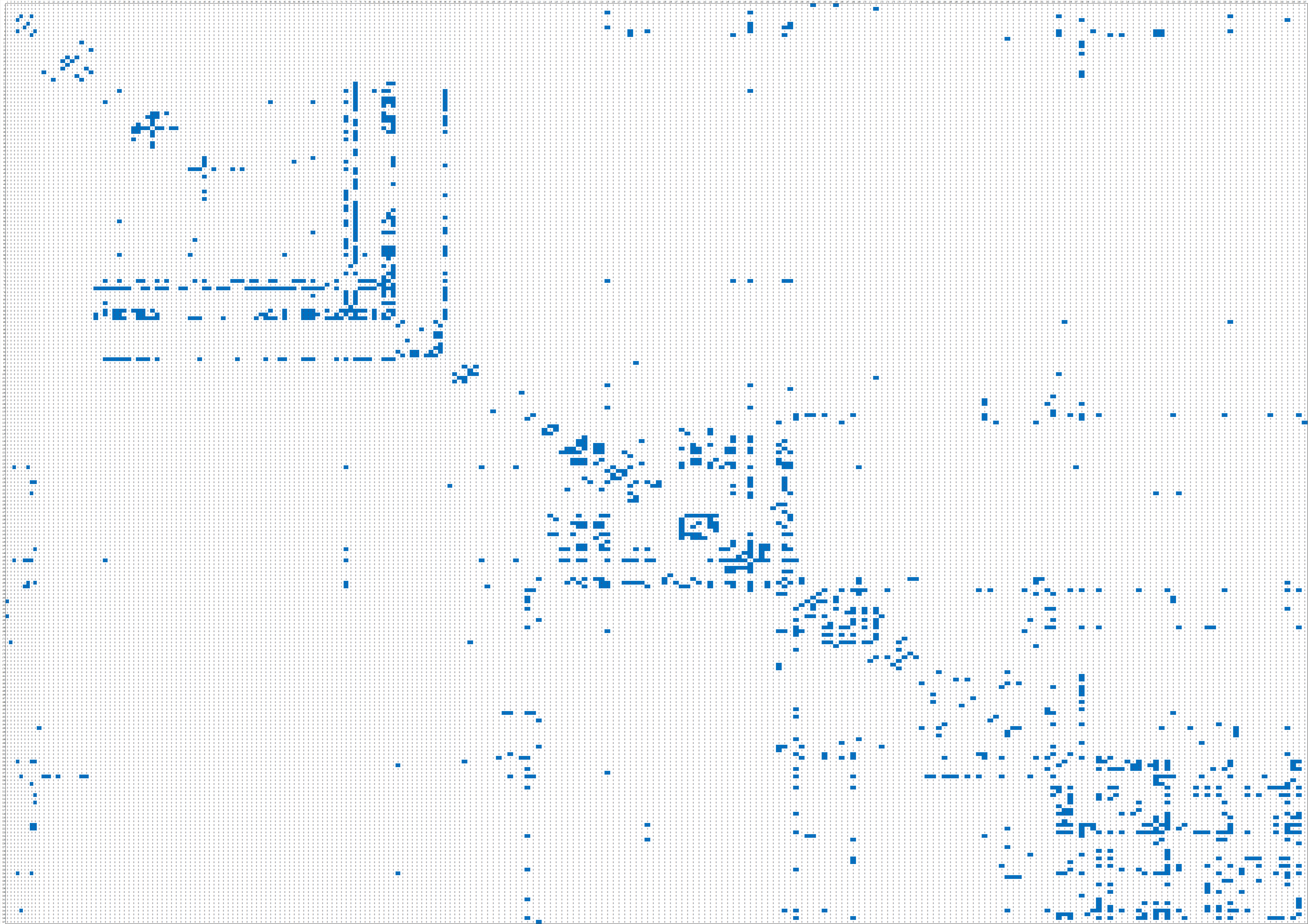}
\end{center}
% \vskip\baselineskip
\subsubsection*{Software Space Proximity Matrix (Dataset 2)}
\begin{center}
\includegraphics[width=1.58\columnwidth]{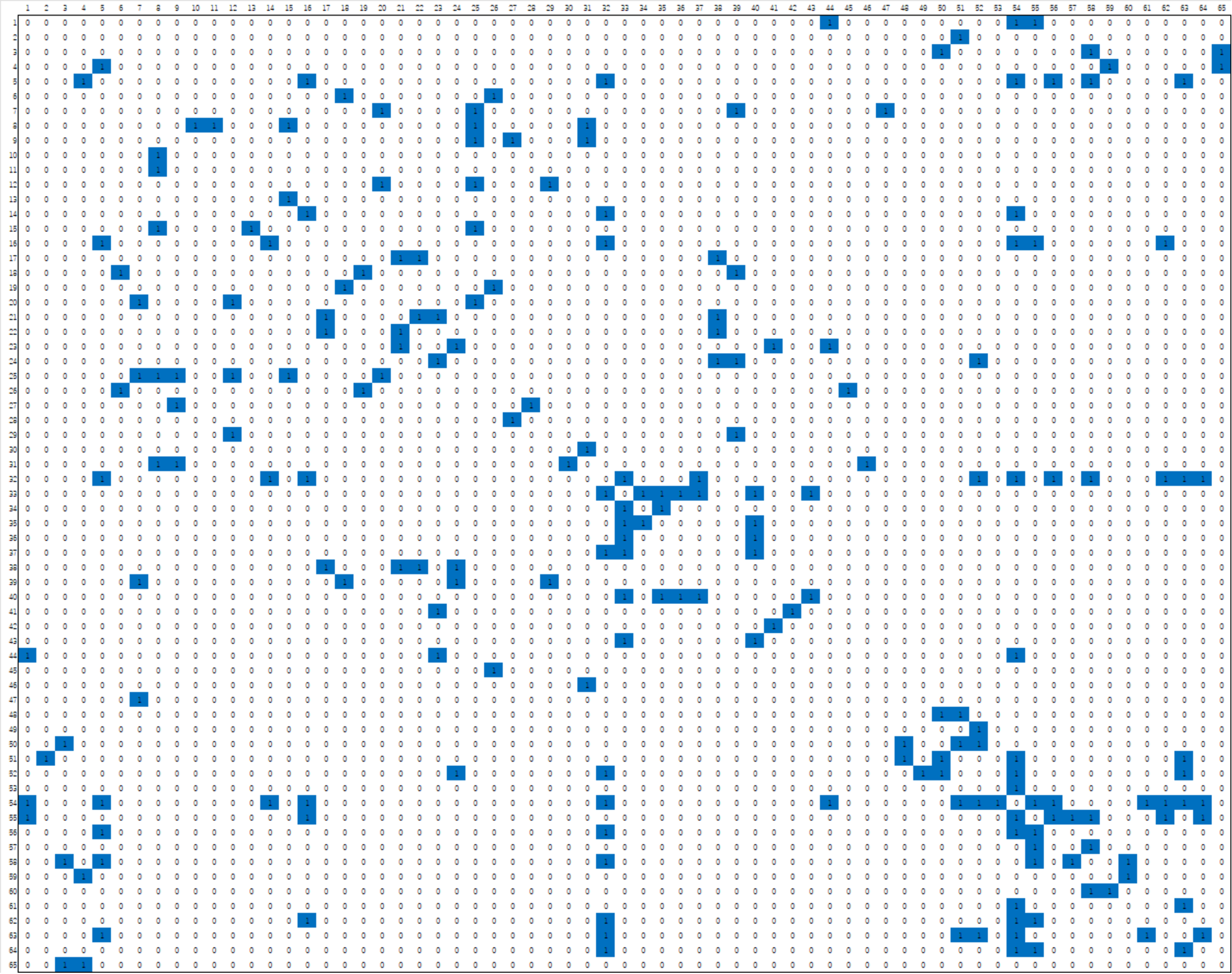}
\end{center}
\clearpage

\subsubsection*{Execution Count Data (Dataset 1)}
\begin{center}
\includegraphics[width=1.52\columnwidth]{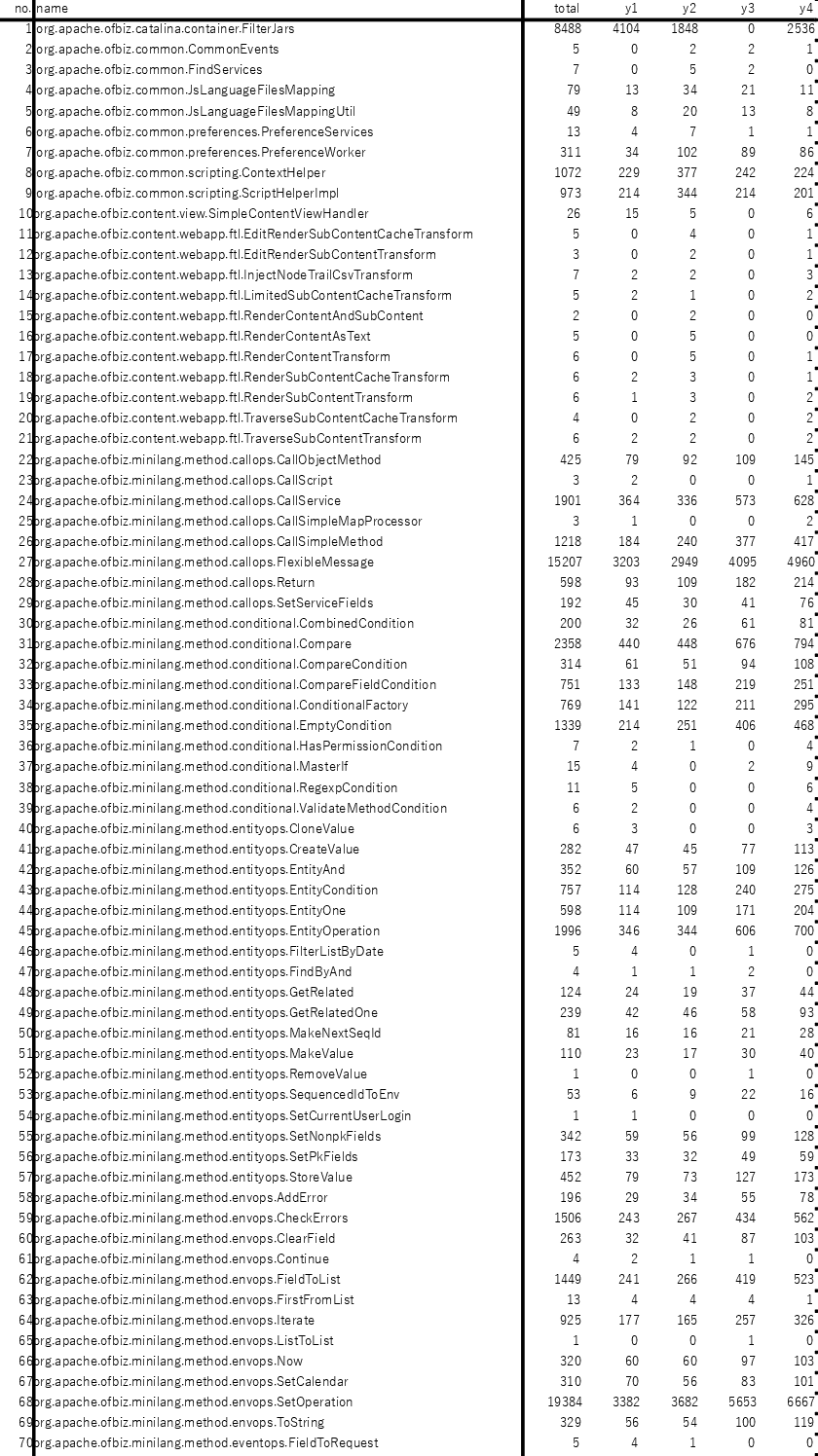}
\end{center}

\clearpage
\begin{center}
\includegraphics[width=1.52\columnwidth]{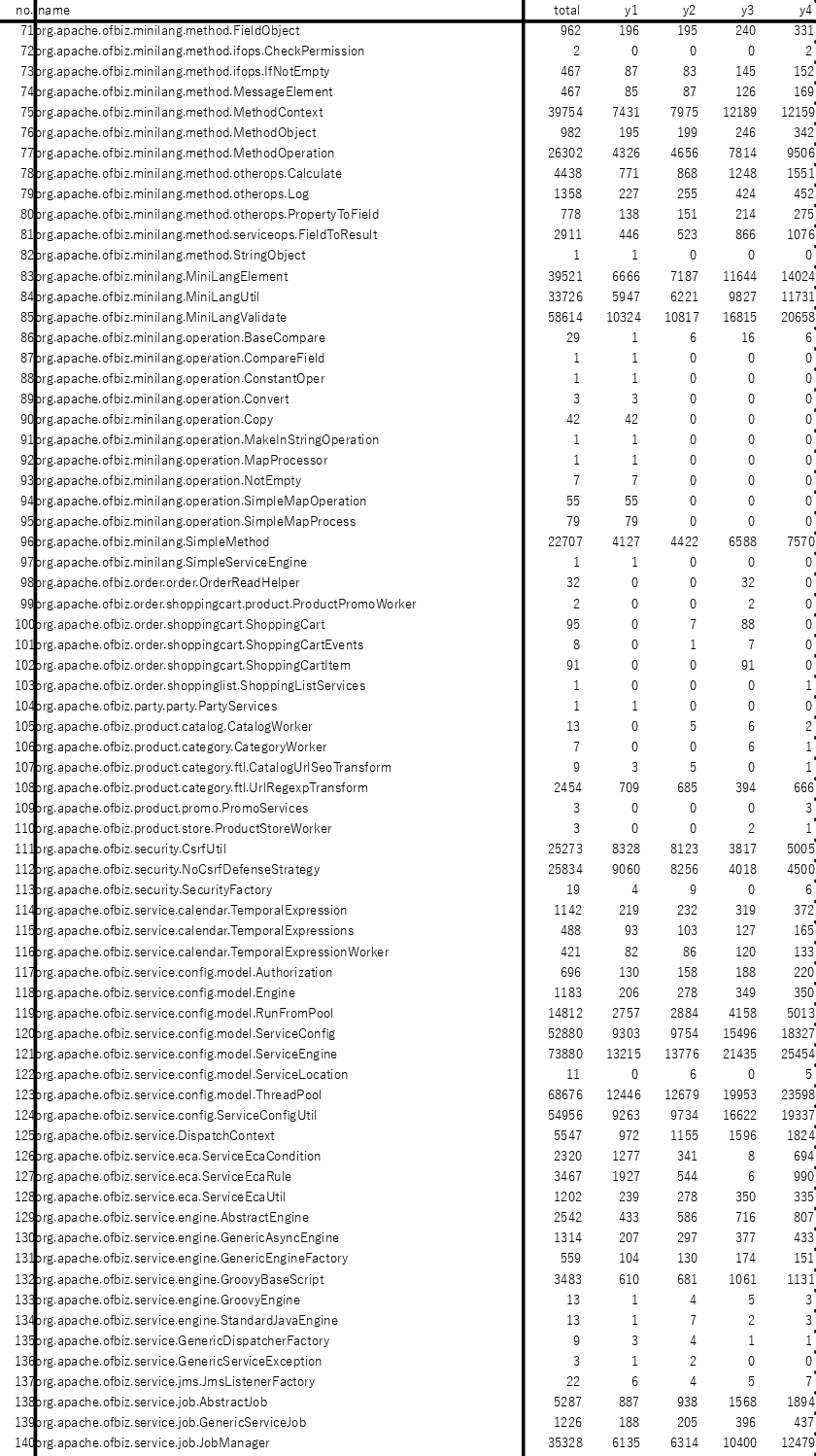}
\end{center}

\clearpage
\begin{center}
\includegraphics[width=1.52\columnwidth]{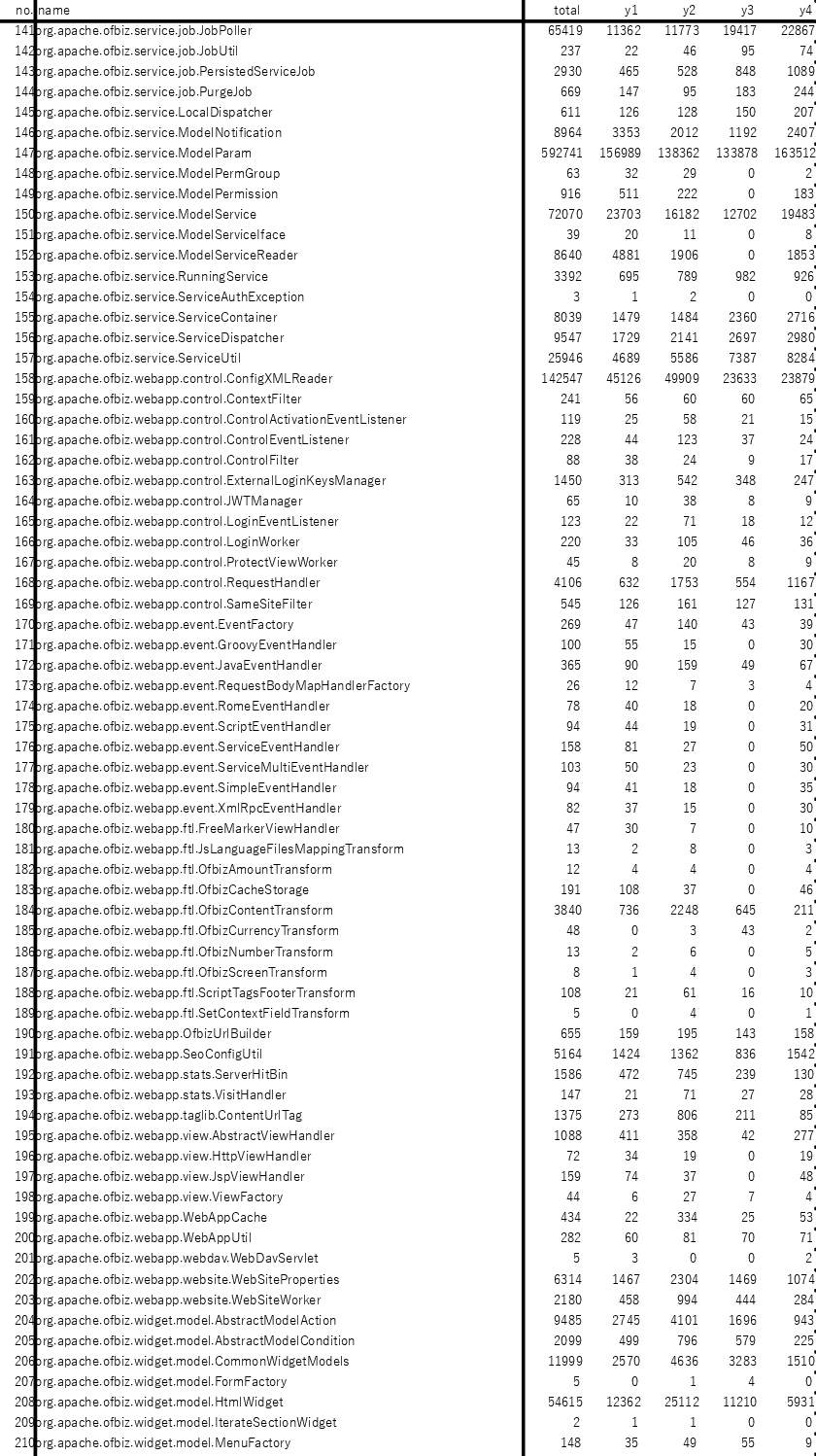}
\end{center}

\clearpage
\begin{center}
\includegraphics[width=1.52\columnwidth]{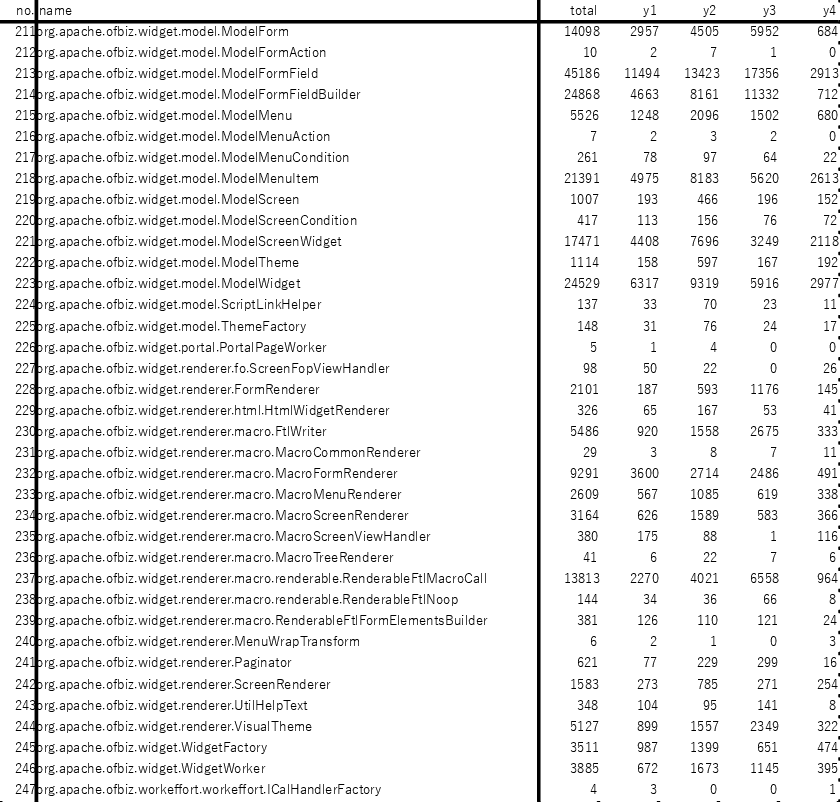}
\end{center}

\clearpage
\subsubsection*{Execution Count Data (Dataset 2)}
\begin{center}
\includegraphics[width=1.52\columnwidth]{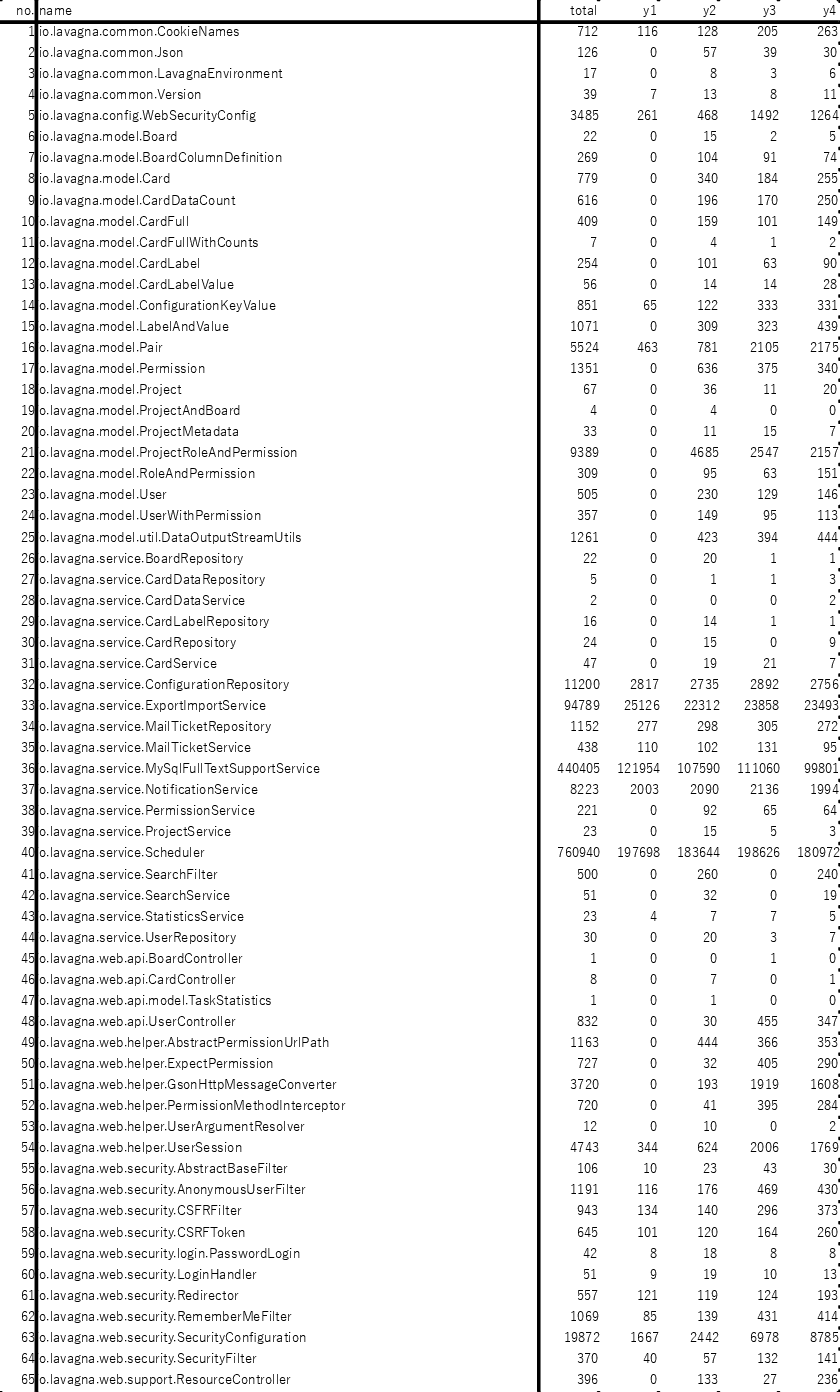}
\end{center}

\end{document}